\documentclass[onecolumn]{aa}
\usepackage{graphicx}
\usepackage{natbib}
\bibpunct{(}{)}{;}{a}{}{,}
\usepackage{txfonts}
\begin{document}

\title{Discovery and photometry  of the binary-lensing caustic-crossing 
event EROS-BLG-2000-5}
\author 
{
C.~Afonso\inst{1,8},
J.N.~Albert\inst{2},
J.~Andersen\inst{6},
R.~Ansari\inst{2},
\'E.~Aubourg\inst{1},
P.~Bareyre\inst{1,4},
G.~Blanc\inst{1},
X.~Charlot\inst{1},
F.~Couchot\inst{2},
C.~Coutures\inst{1},
R.~Ferlet\inst{5},
D.~Fields\inst{7},
P.~Fouqu\'e\inst{9,10},
J.F.~Glicenstein\inst{1},
B.~Goldman\inst{1,8},
A.~Gould\inst{7},
D.~Graff\,\inst{7},
M.~Gros\inst{1},
J.~Ha\"{\i}ssinski\inst{2},
C.~Hamadache\inst{1},
J.~de Kat\inst{1},
L.~LeGuillou\inst{1},
\'E.~Lesquoy\inst{1,5},
C.~Loup\inst{5},
C.~Magneville \inst{1},
J.B.~Marquette\inst{5},
\'E.~Maurice\inst{3},
A.~Maury\inst{9},
A.~Milsztajn \inst{1},
M.~Moniez\inst{2},
N.~Palanque-Delabrouille\inst{1},
O.~Perdereau\inst{2},
L.~Pr\'evot\inst{3},
Y.~Rahal\inst{2},
J.~Rich\inst{1},
M.~Spiro\inst{1},
P.~Tisserand\inst{1},
A.~Vidal-Madjar\inst{5},
L.~Vigroux\inst{1},
S.~Zylberajch\inst{1}
\\   \indent   \indent
The EROS collaboration
}
\offprints{J-F. Glicenstein}

   \institute{
CEA, DSM, DAPNIA,
Centre d'\'Etudes de Saclay, F-91191 Gif-sur-Yvette Cedex, France
\and
Laboratoire de l'Acc\'{e}l\'{e}rateur Lin\'{e}aire,
IN2P3 CNRS, Universit\'e de Paris-Sud, F-91405 Orsay Cedex, France
\and
Observatoire de Marseille,
2 pl. Le Verrier, F-13248 Marseille Cedex 04, France
\and
Coll\`ege de France, Physique Corpusculaire et Cosmologie, IN2P3 CNRS,
11 pl. M. Berthelot, F-75231 Paris Cedex, France
\and
Institut d'Astrophysique de Paris, INSU CNRS,
98~bis Boulevard Arago, F-75014 Paris, France
\and
Astronomical Observatory, Copenhagen University, Juliane Maries Vej 30,
DK-2100 Copenhagen, Denmark
\and
Departments of Astronomy and Physics, Ohio State University, Columbus,
OH 43210, U.S.A.
\and
Department of Astronomy, New Mexico State University, Las Cruces, NM 88003-8001, U.S.A.
\and
European Southern Observatory, Casilla 19001, Santiago 19, Chile
\and
Observatoire de Paris-Meudon LESIA, F92195 Meudon CEDEX, France
              }
\titlerunning{Discovery and photometry of EROS-BLG-2000-5}
\authorrunning{EROS Collaboration}

   \date{Received , accepted }

\abstract{
The EROS collaboration has developed a real time alert system, 
which was operational in 1999 and 2000. A very long
binary-lensing caustic crossing event, EROS-BLG-2000-5, was found 
in the 2000 season and monitored by EROS. EROS photometric 
data for this event are presented. The data are well fitted by  
the model of \citet{An2002}  based on a completely different dataset.
 However, parameters of the model such as the mass ratio or the binary rotation rate are not very well constrained by EROS data. 

} 

	\maketitle

\section{Introduction} 

	Microlensing is becoming established as a powerful tool  in
astrophysics. Astrophysical information on the source can often be obtained 
even with only limited information on the geometry of the lens.  For instance, 
the spatial structure of the atmosphere can be studied when a lens caustic 
passes over the face of a source. For point lenses, the caustics are 
point-like, so that the probability of such a crossing is small. 
For binary lenses, the caustics form one to three polygon-shaped curves 
and the probability of crossing is of the order a few percent. 
Caustics are generic 
singularities of the lens mapping (``catastrophes''). Because of this, the 
shape of the microlensing light curve is accurately described by a limited 
number of parameters (including limb darkening parameters), almost 
independently of the lens model. Intensive photometric observations of 
four binary caustic crossing events have yielded limb-darkening measurements 
\citep{Afonso2000,Albrow1999, Albrow2000,Albrow2001a}.

 The implementation of alert systems by all the major microlensing surveys: 
OGLE \citep{OGLEtrigger}, MACHO 
\citep{MACHOtrigger}, EROS and the
MOA collaboration \citep{MOA} was the key improvement. 
These collaborations  
issued hundreds of microlensing alerts,
found mostly towards the Galactic Bulge. The EROS microlensing alert system,
 which is described in the first part of this paper,
was mainly operational during the 1999 and the 2000 seasons. 
An interesting binary-lensing caustic crossing event, 
EROS-BLG-2000-5, was found during the 2000 season. 
This event was intensively monitored by various  ``follow-up'' collaborations 
such as  PLANET\footnote{http://thales.astro.rug.nl/~planet/}
 and MPS\footnote{http://bustard.phys.nd.edu/MPS/index.html}
and also  by EROS. High resolution spectra 
were taken on the Keck during the second caustic crossing by  
\citet{Castro2001}. Lower resolution spectra were obtained on the VLT by 
\citet{Albrow2001b}. These groups detected a significant
evolution of the equivalent width of the H${\alpha}$ line during the crossing.
\citet{Afonso2001} reanalyzed the spectral data using 
a subsample of EROS photometric data. They found evidence for chromospheric
activity of the K3 giant source. A detailed lens model of EROS-BLG-2000-5 was 
published by \citet{An2002}. The measurement of both
 the finite size of the source and the effect of Earth motion
(``parallax'') allowed the simultaneous determination of the lens distance,  
transverse velocity and mass. 
 An and co-authors
studied limb darkening only for the I passband. However, 
the second caustic crossing of EROS-BLG-2000-5 
was monitored in 3  passbands by PLANET and 2 other passbands by EROS. 
This allows a precise comparison between the observed limb darkening and
the prediction of various atmospheric models \citep{Fields2003}. 
The extraction of limb darkening parameters relies on the agreement 
between the lens model of \citet{An2002} and PLANET data. The photometric data 
reduction for  EROS-BLG-2000-5, and the comparison to the lens model 
of \citet{An2002} is reported in the second part of the paper.

\section{The EROS survey and alert system}

The EROS project used a dedicated 1 meter telescope and 
two 1 $\mbox{deg}^{2}$  cameras at the La Silla ESO site to monitor fields 
in the Magellanic Clouds, the Galactic Bulge and the Galactic Disc. 
The detector setup is described in detail in  \citet{Bauer96} and the data 
acquisition system in \citet{Glicenstein98}. Sixty to eighty images were 
taken every night and flat-fielded online. 
	EROS observations were carried out in two bands $V_E$ and $I_E$. 
 EROS $V_E$ is centered midway between Johnson $V$ and Cousins $R$, while
EROS $I_E$ is similar to Cousins $I$, but broader.
More precisely
\begin{equation}
V_E = 0.69\,V + 0.31\,I ,\quad
I_E = -0.01\,V + 1.01\,I,
\label{eqn:vitrans}
\end{equation} 
 \citep{Regnault2000}.
The disc capacity of the acquisition computers was sufficient 
for storing roughly one night of data taking. 
The data were then copied to tapes and sent to Lyon (France)  for offline 
analysis at the high energy physics computing centre (CC-IN2P3).  
Two dedicated computers were installed on site to perform real-time 
photometry on a limited set of EROS fields. This set included all the EROS 
Small Magellanic Cloud (SMC) fields ($\sim\ 9\ \mbox{deg}^{2}$) 
and roughly one third ($\sim 22\ \mbox{deg}^{2}\ $) of the Galactic 
Bulge (BLG) fields.
Five million stars were thus monitored towards the SMC. Two nov\ae\ were 
discovered and reported \citep{IAU99a,IAU99b}. Only 4.3 million bright 
stars  were  monitored  towards the Galactic Bulge. This bright star 
sample was selected
to minimize the blending of source stars and to disentangle the various 
(disc, bulge) contributions to the microlensing optical depth towards the 
Galactic Centre \citep{Gould1995,Afonso2003}.   
The sampling rate for these fields ranges from one image every 
1-4 nights for the SMC fields
to one image every night for some of the BLG fields. 

The online photometry and alert system is controlled 
by Perl scripts. Each of the scripts in turn 
calls C++ programs written with the PEIDA++ class library. 
The system operation can be split into four basic tasks.    

The first task starts at the end of every night of data taking
and makes lists of images to be processed by the photometry (second) task. 
The latter is similar to the offline photometric reduction. 
It uses our standard PEIDA \citep{Ansari1996} 
PSF-based software. The star catalogues are copies of 
the offline catalogues. The output
of the photometric reduction had to be simplified to save disc space. 
Since we had a limited disc storage capacity on site, the photometric 
reduction had to run in less than a few hours. 
The time needed to process a full image of 8 CCD
was of the order of 15 minutes (both colours running in parallel). 
The number of alert fields was restricted to $\sim 40$ fields 
both by the disc capacity and  the available CPU power.      

The third task is the detection of potential ongoing microlensing events 
(``trigger''). This was done by comparing the template flux of 
each star in the catalogue
to the measured flux. The template fluxes and the typical photometric 
errors were computed offline using $\sim 60$ measurements taken during the 
first 2 years of data taking. 
The ``trigger'' task first builds the light curve in $I_{E}$ and $V_{E}$
of each star in the catalogue. This light curve contains only  data 
points from the running season. Quality cuts are then applied to remove 
data points with large absorption, seeing or skylight. A ``trigger condition''
arises when a star shows a significant deviation from its template flux
simultaneously in both passbands. The distribution of deviations from
the template flux is not distributed as a gaussian. Relatively large 
deviations may occur because of systematic errors in the photometry
(e.g. small errors in the position of sources in the catalogue) or
poor data taking conditions (e.g. large airmasses). Various ``trigger
conditions'' have been tried. Examples are 4 points in a row at more than 
4 times the photometric error, 1 point in both colors at more than 7 times
the photometric error and a minimum increase in flux. The trigger task logs
the information (position, flux, trigger conditions) of stars that satisfy at
least one trigger condition to a file.  
 
The last task builds the light curves of all the candidate 
microlensing events. These candidates are
scanned by eye to remove obvious photometric problems (such as noisy pixels 
or contamination by the light of a nearby luminous star) and finally are run 
through more sophisticated tests. 
The colour of the source is compared to the position of the red giant clump
on the local colour-magnitude diagram. 
Sources redder than the red giant clump, which often show long term
time variations which may fake microlensing are rejected.
Finally, the light curve is fitted to a 
``point lens-point source'' (PLPS) lens model. 
The latter \citep{Paczynski86} is defined by:
\begin{eqnarray}
F^{p}(t)& = A(t) F^{p}_{s} +F^{p}_{b} \\
A &= \frac{u^{2}+2}{u\sqrt{u^{2}+4}} \\
u^{2} &= u_{o}^{2} + {(\frac{t-t_{o}}{t_{E}})}^2
\end{eqnarray}
where $F^{p}, F^{p}_{s} , F^{p}_{b}$ are the total, source and background fluxes in passband 
p, measured at time $t$ and  $t_{E}$ is the Einstein radius crossing time.
The maximum amplification $A_{max}$ of the background source occurs 
at time $t_{o}$. Light curves with $A_{max} < 1.3$ are
rejected.   

An alert is sent to a mail distribution list when the candidate 
passes all the tests. 
Information on the event such as the  coordinates of the source, 
finding charts, colour-magnitude 
diagrams  and parameters of the PLPS fit are displayed on a Web 
page~\footnote{http://www-dapnia.cea.fr/Spp/Experiences/EROS/alertes.html}.
Fifteen microlensing alerts were sent during the 1999 and the 2000 seasons. 
Seven other events were found in 1998 and 2001. 
 
EROS BLG data of the 1996,1997 and 1998 seasons have been analyzed by
\citet{Afonso2003}.
These authors found sixteen microlensing events with red giant sources
and a maximum amplification $A_{max} > 1.34.$
 The four microlensing alerts of the 1998 season were also found 
by \citet{Afonso2003}. However, one of these alerts 
had $A_{max} = 1.32$ and 
thus was not included in the  final offline sample. 
Roughly 20\% of the alerts given by the EROS alert system in 1999 and 2000 
were probable variable stars. Accounting for the smaller efficiency 
of data taking during the first years of EROS2, 
the rate of  online microlensing events detection is typically half 
of the offline detection rate \citep{Afonso2003}.    

\section{EROS-BLG-2000-5}
  This event was alerted by EROS  on 5 May 2000.  
On 8 June 2000, MPS issued an anomaly  alert, informing that a caustic 
crossing was in progress.  Subsequent intensive observations by PLANET
allowed them to predict the time of the second caustic
crossing and, very importantly, that this crossing would last an 
unusually long 4 days. The event was densely monitored by EROS. 
EROS observations are normally carried out in survey mode,
but owing to the importance of the event, extra time was allocated to it. 
Between the caustic crossings, images were taken whenever possible, even on 
cloudy nights ot nights affected by moon light. During the second crossing,
all available time was allocated to it.

\subsection{Photometric Data}

\begin{figure*}
\centering
\includegraphics[width=17cm]{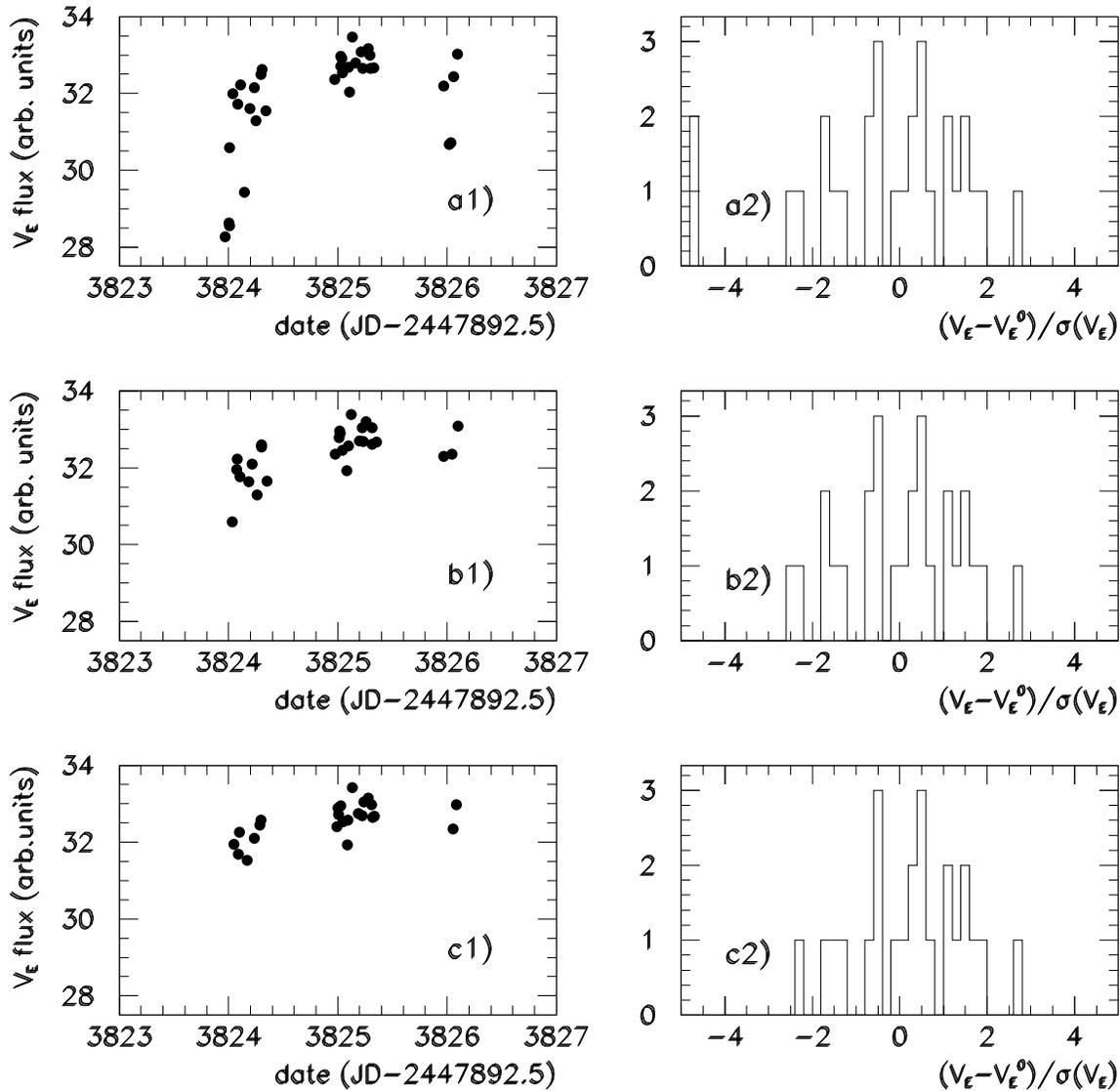}
\caption{
EROS $V_{E}$ data between JD 2451715.5 and JD 2451719.5. 
Plots on the left panel show the light curves and plots on the right panel
show the photometric dispersion. The quality
cut is increasingly demanding from top to bottom.  a) all measurements (
34 points) 
b) measurements with  $n_{\mbox{star}} > 180$ (28 points) 
c) measurements with $n_{\mbox{star}}> 250$ (24 points). 
$V^{o}_{E}$ is the average $V_{E}$ flux calculated with 
$n_{\mbox{star}} > 180.$
$\sigma(V_{E})$ is the ISIS photometry error.  
}
\label{fig:quality}
\end{figure*}

The EROS images have been 
reprocessed using the ISIS \citep{Alard2000} image subtraction 
program on a small 300x300 vignette around the source star. 
The online PEIDA photometry was  used to obtain  
the zero point of image subtracted photometry as explained 
in \citet{Afonso2001} (see their figure 4). 
All the images were sent through the photometry pipeline. 
 However,  some images were of low quality and the photometry showed a 
large dispersion (figure \ref{fig:quality} a). 
The dispersion was reduced by using a quality factor. This quality 
factor is simply the 
 number of stars $n_{\mbox{star}}$ reconstructed 
on the vignette around the source star. 
The effect of $n_{\mbox{star}}$ on the photometric dispersion is 
illustrated on figure \ref{fig:quality}.
Typical values for $n_{\mbox{star}}$ on high quality data range from 
250 to 400 on $V_{E}$ data and from 200 to 250 on $I_{E}$ data.
To avoid losing too many points, somewhat looser cuts were adopted:
$n_{\mbox{star}} > 180$ on $V_{E}$ data and $n_{\mbox{star}} > 170$ on 
$I_{E}$ data.
The cut on $n_{\mbox{stars}}$ is still fairly severe, as shown in table 
\ref{table:data}. Only roughly  two thirds of the points survive the cuts.
The photometric error can be estimated from the dispersion
in the high quality data.
Ideally, the photometric dispersion after the quality cut should be given by 
the photon noise. However, it is found to be slightly larger. The 
true photometric error is obtained from the photon noise dispersion 
by multiplying by the numbers in the fifth column of table 
\ref{table:data}.  

\begin{table}
\centering
\caption[]{EROS data on EROS-BLG-2000-5 }
      $$  \begin{array}{p{0.2\linewidth}c|c|c|c|c|}
            \hline
 \mbox{filter} &  \mbox{images} & \mbox{images after quality cut} &  \mbox{selected for fit} 
& \mbox{error renormalization factor} \\
\hline
$V_{E}$ &  1275 & 1021 & 830 & 1.24  \\
$I_{E}$ &  1207 & 904 & 904 &  1.16 \\
\hline
         \end{array}
$$
\label{table:data}
\end{table}


\subsection{Lens model and discussion}
Our data have been fitted to a variant of the model of 
\citet{An2002}. This model takes into account both the 
finite size of the  source (through the projection of the source star 
on the lens plane $\rho_{\star}$), and the two components 
of projected motion of 
the Earth on the  lens plane (``parallax'') $\pi_{E,||}$ parallel 
and $\pi_{E,\perp}$ perpendicular to the binary axis. The geometry of 
the various projections  is shown in figure 2 of \citet{An2002}.  
The other parameters are the time of closest approach to the cusp $t_{c},$ 
the distance $d_{t_{c}}$ of the 2 components of the binary 
evaluated at $t_{c},$ 
 the time variation of this 
distance $\dot{\vec{d}},$ the mass 
ratio $q,$  the distance 
of closest approach $u_{c},$ the Einstein radius 
crossing time $t'_{E},$ the angle between the lens
trajectory and the binary axis $\alpha'$ and the binary rotation rate 
$\omega.$    
The total number of geometric parameters is thus 11. 
There are also 4 additionnal parameters 
per passband: the source flux, the background flux and 2 limb 
darkening parameters. In the original An et al. model, only I passband 
was used, but the images came from several observatories. Because of this,
a different source flux, background flux and seeing correction had to be
fitted for each observatory.

The fitting procedure is described in detail in  
\citet{An2002} and \citet{Fields2003}. 
The minimum $\chi^{2}$ is found by stepping over a grid of $(d_{t_{c}},q)$ 
values and minimizing over the other parameters. The apparent 
$\chi^{2}$ surface has a very complicated and ``rough'' shape around the
minimum as explained in \citet{Fields2003}. 
Multiple local minima are possible. 
The numerical evaluation of errors is made difficult by the 
roughness of the  $\chi^{2}$  surface.

\begin{figure}
\includegraphics{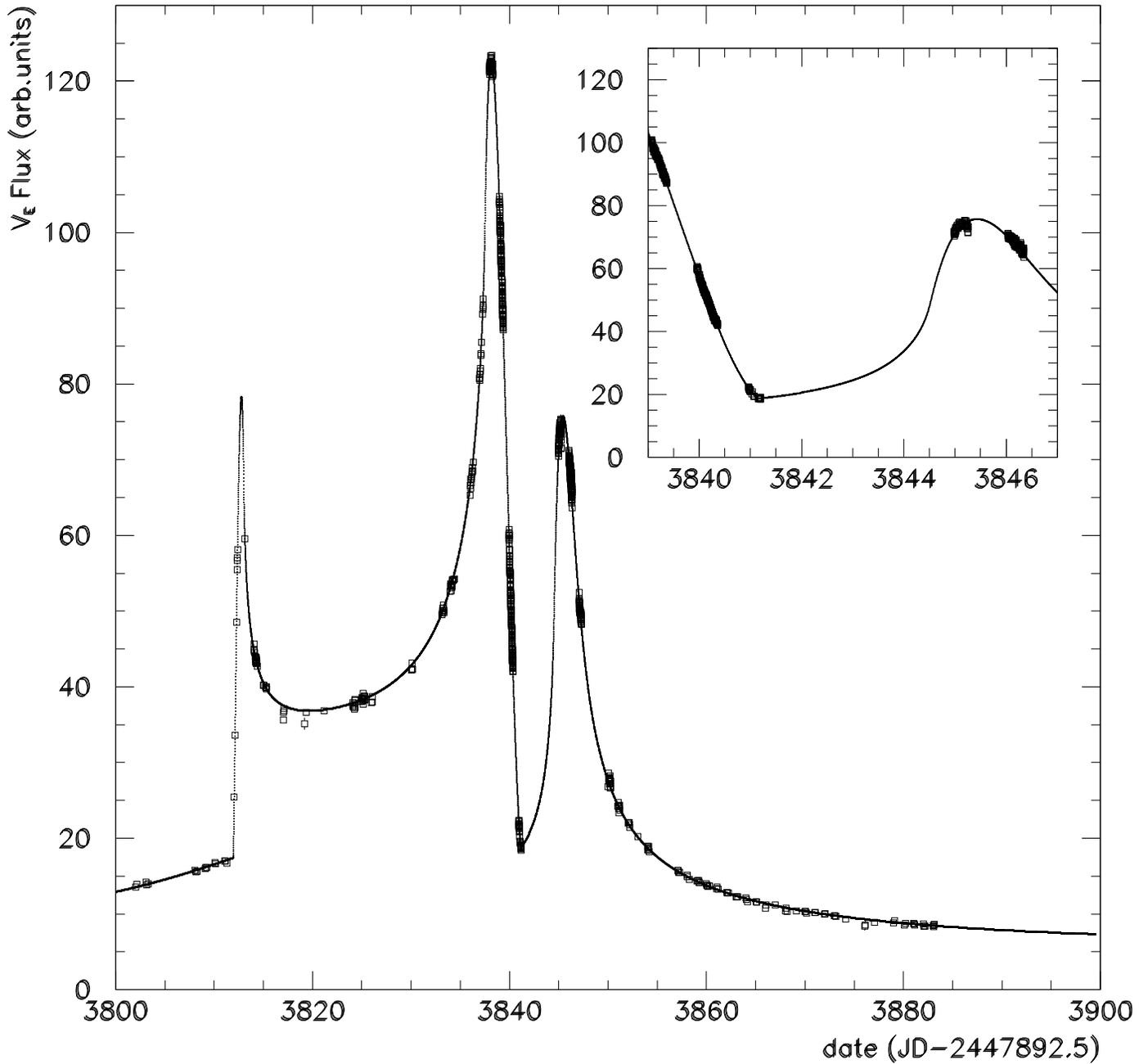}
\caption{
EROS $V_{E}$ data (squares) compared  to the \citet{An2002} lens model (solid line). 
The blowup shows the end of the second caustic crossing and the cusp approach.
}
\label{fig:erosb}
\end{figure}

\begin{figure}
\includegraphics{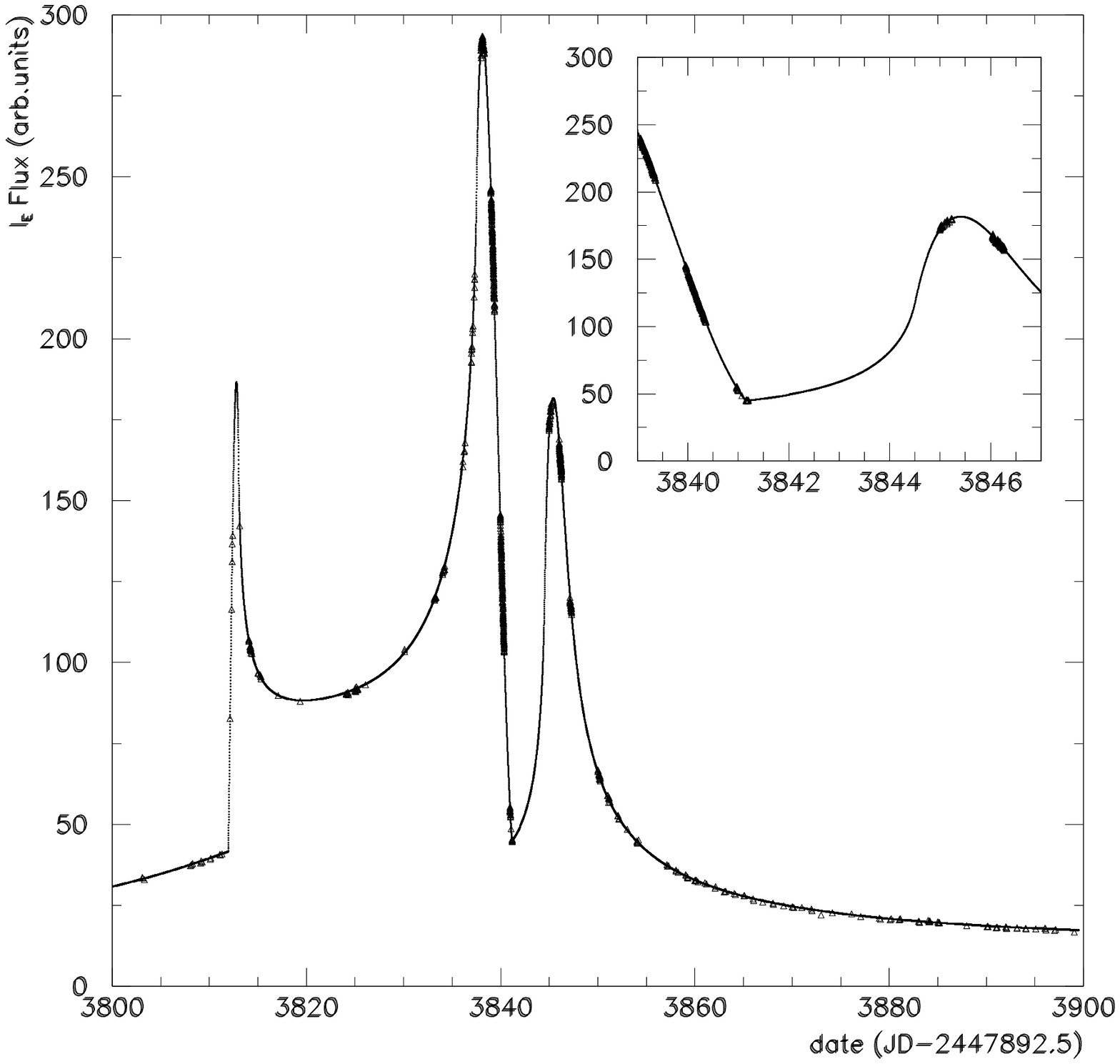}
\caption{
EROS $I_{E}$ data (triangles) compared  to the \citet{An2002} lens model (solid line).
The blowup shows the end of the second caustic crossing and the cusp approach.
}
\label{fig:erosr}
\end{figure}

\begin{figure}
\includegraphics{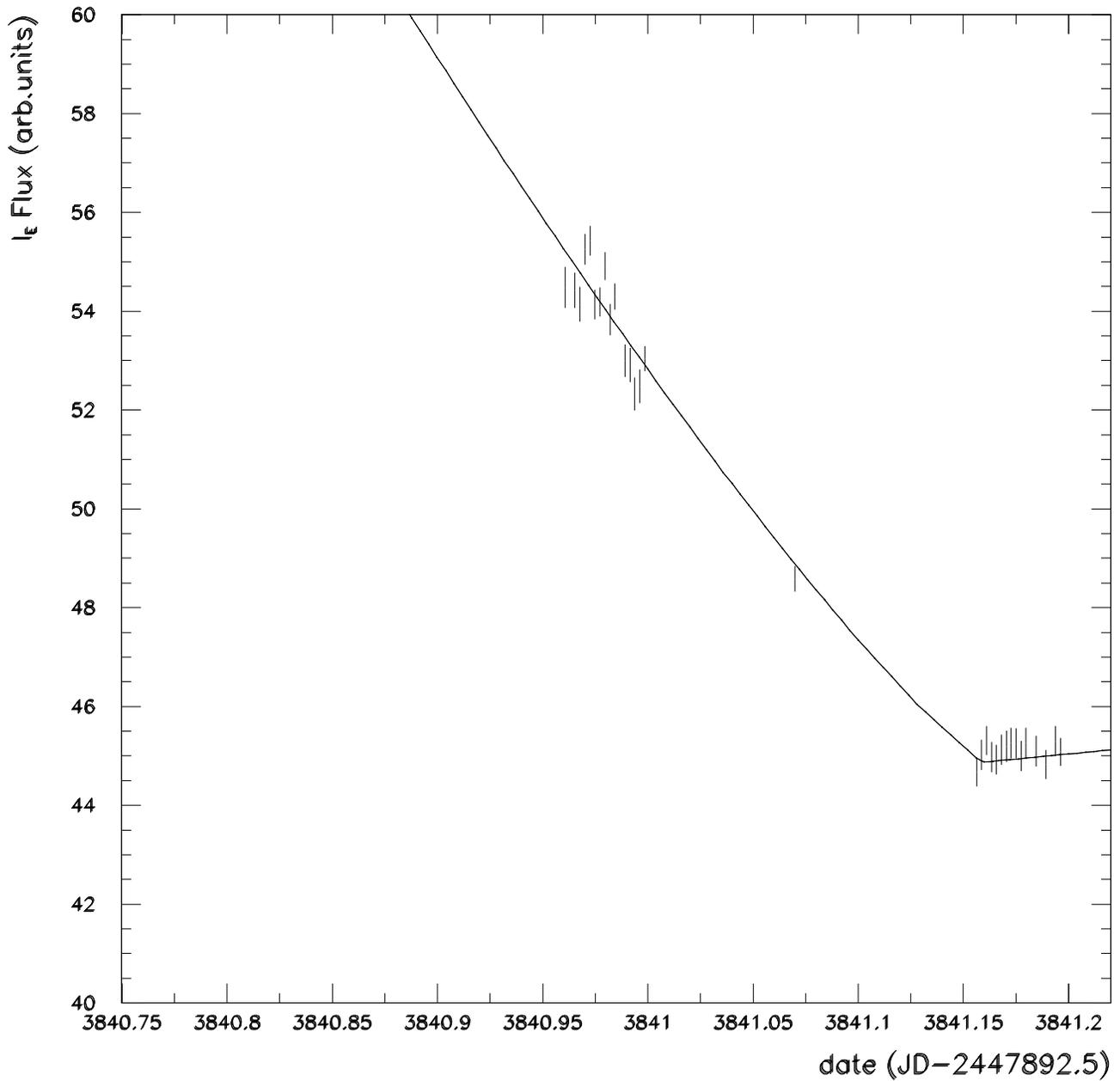}
\caption{
Blowup of the exit of second caustic crossing for 
EROS $I_{E}$ data. The solid line is the prediction of \citet{An2002}. Note that this model is based on a dataset where the caustic exit was not observed.
}
\label{fig:erosblowup}
\end{figure}

EROS $V_{E}$ and $I_{E}$  data are displayed in figures \ref{fig:erosb} and 
\ref{fig:erosr}. The \citet{An2002} lens model, which is 
superimposed on the data, gives obviously an excellent fit. 
The location of the second caustic exit is clearly seen in EROS data 
(see figure \ref{fig:erosblowup}) and is in very good agreement with the prediction of the model. Note that this prediction was not tested in the \citet{An2002} paper because of the lack of data immediately after the end of the caustic crossing (see their section 3.2).   
 
However, several other $(d_{t_{c}},q)$ values are found to give excellent fits 
to the EROS data. Two examples of such fits (local minima of the apparent
$\chi^{2}$ surface), labelled ``Model 1'' 
and ``Model 2'' are shown in table 
\ref{table:fit} and compared to the \citet{An2002} values. 
\begin{table}
\centering
\caption[]{Parameters of lens models for EROS-BLG-2000-5. The fit errors were not calculated for models 1 and 2 because of numerical problems 
(see \citet{Fields2003}).
The last two columns are taken from \citet{An2002} table 3. }
$$       \begin{array}{p{0.2\linewidth}cccc}
            \hline
            \noalign{\smallskip}
\ & \mbox{Model 1} & \mbox{Model 2} & \mbox{An et al} & \  \\
\mbox{Parameter} &\mbox{Value}& \mbox{Value}&  \mbox{Value}& \mbox{Uncertainty}\\
\noalign{\smallskip}
\hline
$d_{t_{c}}$ & 1.935 & 1.935  &1.928 & 0.004\\
$q$& 0.77 & 0.76 &0.7485 & 0.0066\\
$\alpha'(deg)$& 73.82  & 73.62 &74.18  &0.41 \\ 
$u_{c}$ & -5.11 {\cdot} 10^{-3}  & -5.12 \cdot 10^{-3} & -5.2 {\cdot}
10^{-3}&
3 \cdot 10^{-5}\\
$t'_{E}$ (days) &100.9 &  99.7 & 99.8 & 1.5\\
$t_{c}$ (days) &3844.439 & 3844.438 & 3844.444 & 0.005\\
$\rho_{\star}$ &4.88 \cdot 10^{-3} &4.92 \cdot 10^{-3} & 4.80 \cdot
10^{-3}& 4 \cdot 10^{-5}\\
$\pi_{E,||}$ &-0.126 &-0.164 & -0.165& 0.042\\
$\pi_{E,\perp}$&0.169  & 0.148 & 0.222& 0.031\\
$\dot{d} (yr^{-1})$&0.24  & 0.31 & 0.203& 0.016\\
$\omega (yr^{-1})$ & -0.05  &0.019  & 0.006 & 0.076\\
           \hline
$\chi^{2}/d.o.f$ &1712.1/(1734-19) &1713.8/(1734-19)& 2255/(1734-19)& \ \\
         \end{array}
$$
\label{table:fit}
\end{table}

\begin{figure}
\includegraphics{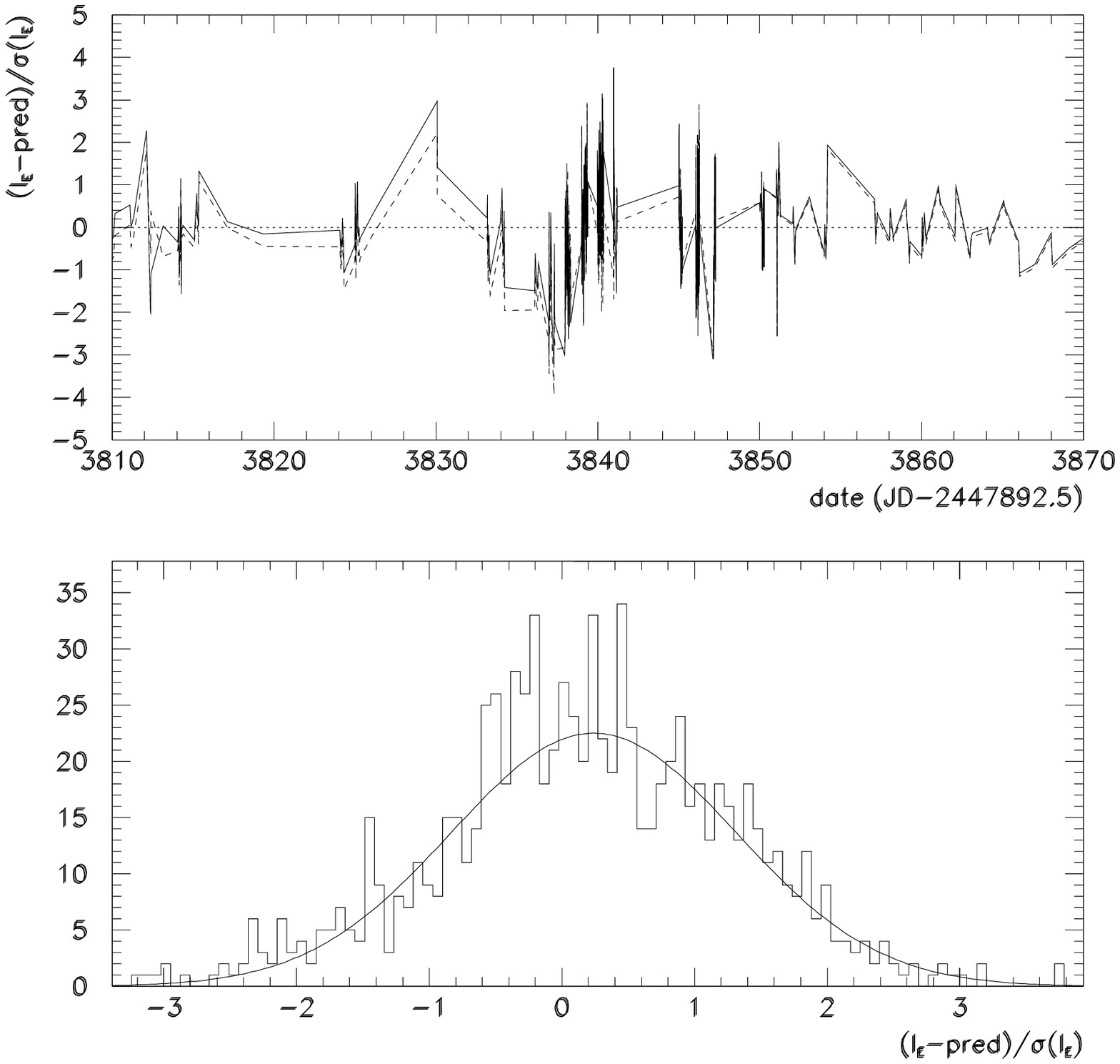}
\caption{
Normalized residuals of EROS $I_{E}$ data. Top: residuals 
as a function of time. The solid line is model 1
(model2 is not shown but similar) and the dashed line is 
\citet{An2002} lens model.
The residuals of both models do not show any obvious trend as a function of time. Bottom: distribution of residuals for model 1. 
The solid line shows the fit to a gaussian.
}\label{fig:erosrres}
\end{figure}

\begin{figure}
\includegraphics{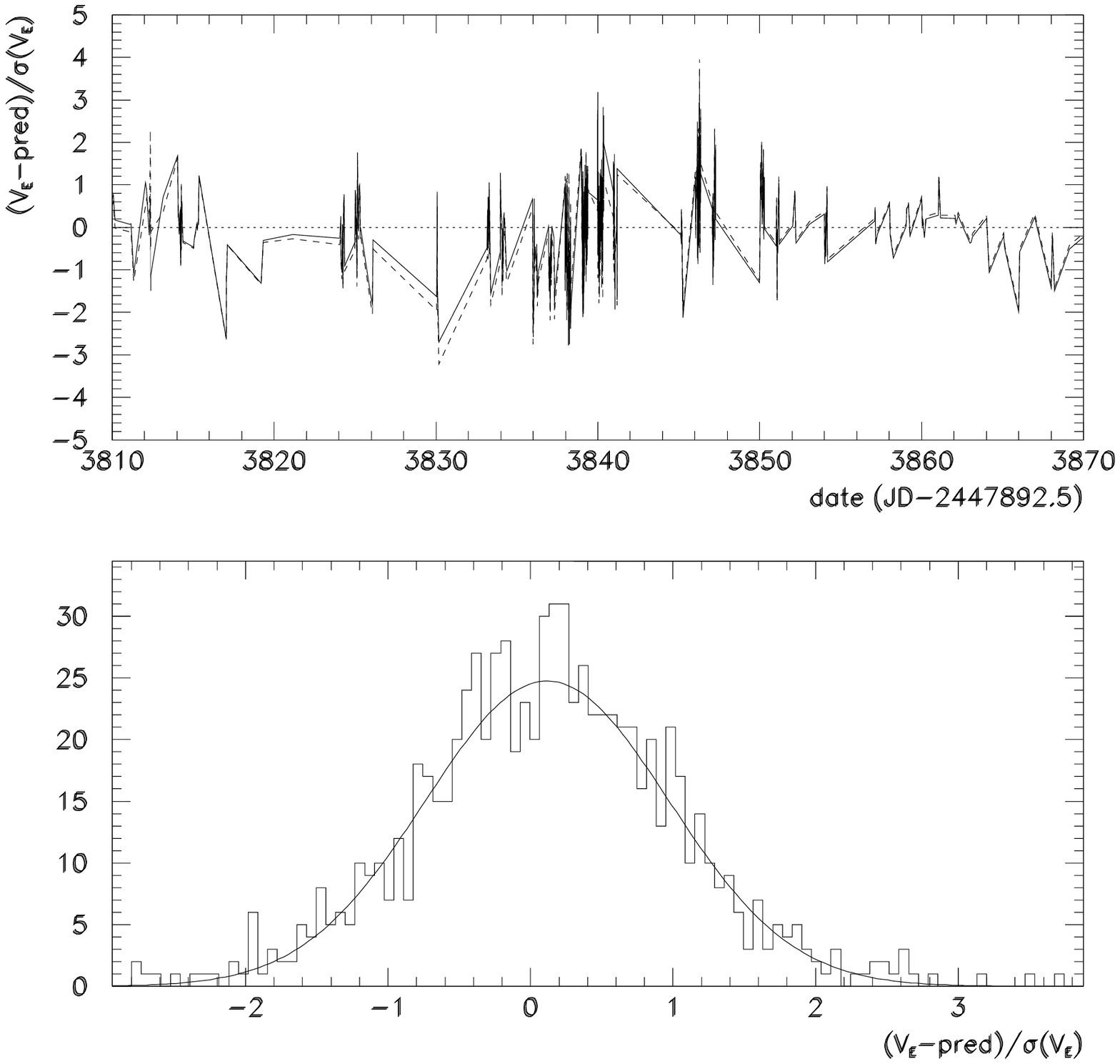}
\caption{
Normalized residuals of EROS $V_{E}$ data. 
Top: residuals 
as a function of time. The solid line is model 1 (model 2 is not shown, 
but similar) and the dashed line is 
\citet{An2002} lens model.
The residuals of both models do not show any obvious trend as a function of 
time. 
Bottom: distribution of residuals for model 1. The line shows the fit to a 
gaussian.
}
\label{fig:erosbres}
\end{figure}

The residuals of the fit of model 1, normalized by the photometric errors,
 are shown in figures
\ref{fig:erosrres} and \ref{fig:erosbres}. These residuals do not show  
any systematic trend as a function of time. The distributions 
of residuals $r_{V_{E}}$ and $r_{I_{E}}$ are well fitted 
by gaussian distributions in both $I_{E}$ and $V_{E}$ filters.
The widths $\sigma(r_{V_{E}})$ and  $\sigma(r_{I_{E}})$ of the 
distributions are close to unity:

\begin{eqnarray}
\sigma(r_{V_{E}}) &= 0.86  \pm 0.03 \\
\sigma(r_{I_{E}}) &= 1.07 \pm 0.03
\end{eqnarray}
  
The prediction of model 1 is systematically smaller than the data in 
both passbands,
but the difference is only $\sim 1/10$ of the typical error bar. 
It is clear from table \ref{table:fit} that 
some parameters are not very well  constrained by EROS data. The parameters 
$q$ and $\pi_{E,\perp}$
are mildly inconsistent with the values found by \citet{An2002}, 
while the value of  $\dot{d}$ in model 2 is highly inconsistent.

\section{Summary}
The binary-lensing caustic-crossing microlensing event EROS-BLG-2000-5 was found by the EROS alert system during the 2000 Bulge season. The EROS data are 
well fitted by the lens model of \citet{An2002}. In particular, this model
predicts the observed end of the second caustic crossing.
However, EROS data are also consistent with other slightly different models. 
The combined fit of all EROS and PLANET  data \citep{Fields2003} should 
hopefully improve the situation.

 \begin{acknowledgements}We are grateful to D.\ Lacroix and the technical staff at the Observatoire de
Haute Provence and to A.\ Baranne for their help with the MARLY
telescope.  We are also grateful for the support
given to our project by the technical staff at ESO, La Silla. We thank
J.F.\ Lecointe and A.\ Gomes for assistance with the online computing.
Work by AG and DF was supported by NSF grant AST~02-01266.
\end{acknowledgements}

\bibliographystyle{aa} 
\bibliography{eros2k5}
\clearpage

\end{document}